\documentclass[trackchanges,onecolumn]{aastex701}

\newcommand\offset{$124 \ \mathrm{pc}$}

\newcommand\ellipsemajor{$0\farcs020$}
\newcommand\ellipseminor{$0\farcs010$}

\newcommand{\chandra}{\mbox{\em Chandra\/}}
\newcommand{\gaia}{\mbox{\em Gaia\/}}

\newcommand{\xmm}{{\it XMM}}
\newcommand{\xmmnew}{{\it XMM-{\it Newton}}}

\newcommand{\gravlens}{\mbox{\textsc{gravlens}\/}}

\newcommand{\euclid}{\mbox{\em Euclid\/}}

\newcommand{\saotrace}{\mbox{\textsc{SAOTrace}\/}}

\newcommand{\lisa}{\mbox{\em LISA\/}}

\newcommand{\rubin}{\mbox{\em Rubin\/}}

   \usepackage{xcolor}
   
\newcommand{\ciao}{\mbox{\textsc{CIAO}\/}}
\newcommand{\xspec}{\mbox{\textsc{Xspec}\/}}
\newcommand{\sherpa}{\mbox{\textsc{Sherpa}\/}}
\newcommand{\marx}
{\mbox{\textsc{MARX}\/}}
\newcommand{\baymax}{\mbox{\textsc{Baymax}\/}}
\newcommand{\hst}{\mbox{\em HST\/}}
\newcommand{\keck}{\mbox{\em Keck\/}}

\newcommand{\axis}{\mbox{\em AXIS\/}}
\newcommand{\hubble}{\mbox{\em Hubble\/}}

\usepackage{ulem}

\usepackage{natbib}
\begin{document}

\title{Determining the spatial origin of X-ray and optical emission in the $z = 3.1$ strongly lensed radio-quiet quasar GraL J065904.1+162909 to hundreds of parsecs}

\author[orcid=0000-0003-3814-6796]{J\'ulia M. Sisk-Reyn\'es}
\affiliation{Center for Astrophysics | Harvard \& Smithsonian, Cambridge, MA, 02138}
\email[show]{julia.sisk\_reynes@cfa.harvard.edu}  

\author[orcid=0000-0001-8252-4753]{Daniel A. Schwartz} 
\affiliation{Center for Astrophysics | Harvard \& Smithsonian, Cambridge, MA, 02138}
\email{fakeemail2@google.com}

\author[orcid=0000-0001-5655-4158]{Anna Barnacka}
\affiliation{Center for Astrophysics | Harvard \& Smithsonian, Cambridge, MA, 02138}
\email{fakeemail3@google.com}

\author[orcid=0000-0002-2231-6861]{Cristiana Spingola}
\affiliation{INAF - Istituto di Radioastronomia, Via Gobetti 101, I-40129, Bologna, Italy}
\email{fakeemail3@google.com}
\author[orcid=0000-0003-0216-8053]{Giulia Migliori}
\affiliation{INAF - Istituto di Radioastronomia, Via Gobetti 101, I-40129, Bologna, Italy}
\email{fakeemail4@google.com}
\begin{abstract}

We perform milliarcsecond X-ray astrometry of the quadruply lensed radio-quiet quasar GraL~J065904.1+162909 (J0659).~This $z_\mathrm{s} = 3.083$ quasar is lensed into four images and was discovered with the second Data Release of the \gaia\ Space Observatory (\gaia\ DR2). Our J0659 study exploits strong gravitational lenses as high resolution telescopes. This technique shows promise to elucidate the origin of optical and X-ray emission in distant lensed quasars at spatial scales beyond the reach of current instruments. In our study, we use \gaia\ DR3 and \hst\ observations of J0659 to infer a mass model for the deflector. Our model reproduces the \gaia\ DR3 quasar lensed image positions to one milliarcsecond and determines the position of the optical source in J0659 to within this precision. Next, we analyze \chandra\ observations of J0659 and conduct a Bayesian test evaluating whether the X-ray emission region coincides with the optical source. We then constrain the origin of the X-ray emission to within a \ellipsemajor$\times$\ellipseminor\ ellipse centered $0\farcs014$ away from the optical source at the $1\sigma$ level. We demonstrate that our approach can be extended to pinpoint the distinct origins of the soft and hard X-ray emission regions in lensed quasars. We discuss the potential of upcoming broadband and spectrally resolved X-ray astrometric studies to probe complex quasar morphology and AGN multiplicity at sub-kiloparsec scales otherwise inaccessible at high redshifts.

\end{abstract}


\section{Introduction} 
\label{sec:introduction}

Supermassive black holes (SMBHs) reside at the centers of all massive galaxies. Active galactic nuclei (AGN) are found at the centers of 1-10\% of all massive galaxies and are indicative of a specific phase of SMBH activity. AGN exhibit the most extreme long-lived luminosities in the universe \citep{padovani_2017_review} and are powered by the accretion of matter onto the SMBH from the surrounding accretion disk \citep[][]{lynden-bell_1974_disks,pringle_1981_disks}.~The AGN unification scheme posits that the plethora of observational properties exhibited by AGN \citep{kellermann_1989_radioloud_uni,urry_1995_unification_scheme,padovani_1995_unification_1} are driven by orientation effects to the same underlying geometry as set by the viewing angle to the obscuring torus \citep{antonucci_miller_1985_tori,antonucci_1993_uni_review}.

Whereas the centroids of X-ray and optical emission in AGN are typically associated with processes within tens of gravitational radii of the SMBH, X-ray emission is also linked to AGN outflows, radio jets, and AGN pairs.~These processes are central to our understanding of SMBH and galaxy coevolution and typically occur on $ 10 \ \mathrm{pc} - 1 \ \mathrm{kpc}$ scales at $ z \ll 1$. However, our ability to resolve them at $z \approx 1 - 3$ is limited by the spatial resolution of current instruments. In the X‑ray band, \chandra’s state‑of‑the‑art angular resolution (0\farcs5 on‑axis) enables detailed studies of these processes at $z \ll 1$ \citep{slane_2025_chandra-nature}. Investigations at higher redshifts are expected to remain scarce as no high-energy missions in the near future will exceed \chandra’s angular resolution.

\chandra\ investigations of unlensed, low-redshift AGN underscore the need to resolve sub-kiloparsec spatial offsets in high-redshift AGN. Firstly, the optical nucleus of M87 ($z = 0.004$) is $\sim 60 \ \mathrm{pc}$ offset from the HST‑1 jet knot whose X-ray and gamma-ray emission dominated the broadband emission from this low-luminosity radio galaxy for many years \citep{stawarz_2006_hst1+m87,harris_2006_hst1+m87,harris_2009_hst-1+m87_gamma}. Secondly, \citet{trindade-falcao_2023_ngc5728,trindade-falcao_2024_ngc5728} used spectrally resolved \chandra\ imaging of the type-2 obscured AGN NGC~5728 ($z \approx 0.01$) and located the X-ray centroid $84 \ \mathrm{pc}$ away from the optical core. The authors also detected X-ray emission in directions parallel and perpendicular to the biconical outflow, underlying a non-standard (porous) torus geometry. \citet{trindade-falcao_2024_ngc5728} also detected semi‑relativistic Fe K$\alpha$ emission $\sim$2 kpc away from the nucleus, interpreted as reflection from ionized clouds beyond the torus.~Thirdly, probing AGN multiplicity on scales of $\mathcal{O}(10-100 \ \mathrm{pc})$ is essential for disentangling and identifying offset AGN, and dual or binary AGN. We use the terminology of \citet{burke-spolaor_2018_proc} and use the terms `dual AGN' and `binary AGN' to refer to SMBHs separated by $> 100 \ \mathrm{pc}$ and $< 100 \ \mathrm{pc}$, respectively. Identifying these systems observationally provides critical constraints on galaxy merger rates and underpins forecasts of the gravitational‑wave signals expected from coalescing SMBHs \citep{dorazio+charisi_2023_SMBH-binaries}.

This work continues a series of papers where we exploit the flux amplification and spatial magnification in strong lensing to investigate the X-ray emission in AGN in a range of spatial scales otherwise inaccessible at cosmological distances \citep{GLAD+schwartz_2021_j2019,GLAD+spingola_2022_2systems,GLAD+rogers_2025_he0435}. We use the caustic method, an innovative technique that allows us to resolve the origin of X‑ray emission by improving both \chandra’s spatial resolution and astrometric precision (nominally,~$\sim 0\farcs6 - 0\farcs8$).~This method leverages caustics --loci connecting source positions where flux amplification is maximal-- as nonlinear amplifiers, enabling the localization of the origin of the X‑ray emission relative to the optical emission to milliarcsecond precision. We seek to apply this method to 24 magnification-selected quadruply lensed quasars (including J0659, which this paper focuses on) covering the redshift range $ z = 1.6 - 4$ with existing \chandra\ X-ray and \gaia\ optical data. Our goal is to constrain or determine X-ray and optical offsets and probe AGN multiplicity.

In this paper we use \gaia, \hst, and \chandra\ observations of the radio-quiet lensed quasar J0659 ($z = 3.1$) and conduct its first X-ray-to-optical astrometric analysis. This makes J0659 the fifth quadruply lensed AGN which the caustic method has been applied to.~For the first time,~we also extend the method to demonstrate that it can localize the distinct origins of hard and soft X-ray emission relative to the optical to milliarcsecond precision.

This paper is organized as follows.~Section \ref{sec:methodology} outlines the caustic method.~Section \ref{sec:gaia} introduces previous works on J0659 and our lens mass model inferred from \gaia\ DR3 and \hst\ archival data.~Section \ref{sec:chandra}~outlines our spectral analysis of the archival \chandra~observations of J0659. We then conduct ray-trace simulations of these \chandra\ data. In Section \ref{sec:results},~we determine the origin of the broadband, soft, and hard X-ray emissions in J0659 relative to the optical source to milliarcsecond precision.~Section \ref{sec:discussion}~discusses the implications of our results in the context of the AGN unification model. We also reflect on the prospects of applying our methods to uncover complex AGN morphology at high redshifts, and to identify binary and dual AGN.~We discuss these prospects in context of ongoing and near-future lens discovery engines, including \euclid, \rubin, and \emph{Roman}. We conclude in Section~\ref{sec:conclusions}.

\section{Methodology}
\label{sec:methodology}

The caustic method \citep{barnacka_2018_rev}~leverages \gaia's~exquisite astrometry and \chandra's imaging capabilities, enabling the localization of the optical and X-ray emission regions in J0659 through a double maximum likelihood (ML) approach.~This method was first applied to radio and optical observations of the quadruply lensed systems CLASS B0712+472 and CLASS B1608+656 \citep[][]{GLAD+firstpaper_2020}.~We have subsequently applied it to probe the origins of broadband X-ray (0.5 - 7 keV observed frame) and optical or radio emissions in these two systems~\citep{GLAD+spingola_2022_2systems}, as well as the following quadruply lensed AGN:~J~2019+1127 \citep{GLAD+schwartz_2021_j2019}, HE 0435-1223 \citep{GLAD+rogers_2025_he0435}, and J0659 (this work). We refer to Section 3.3 of \citet{GLAD+schwartz_2021_j2019} and to the Appendix of \citet{GLAD+rogers_2025_he0435} for a detailed description of this method,~summarized below.

For our J0659 study, we first obtain a model for the lensing mass distribution, utilizing the positions of the quasar lensed images reported by \gaia\ DR3, as informed by~\hst\ observations.~We employ the \gaia\ DR3 positions of the four lensed images since \gaia\ is the most precise astrometric mission to date \citep[][]{gaia_1,gaia_2,gaia_3}. Once our mass model reproduces these observed positions to milliarcsecond precision, the achromaticity of strong lensing ensures that the X-ray lensed images will be at the exact same positions as the \gaia\ DR3 lensed images \textit{if and only if} the X-ray and optical sources are co-located. We test this hypothesis using a double maximum likelihood (ML) test. To begin with, we conduct a high fidelity simulation of the \chandra\ telescope and detector response for each of the four X-ray lensed images using \saotrace\ and \marx, respectively. These simulations are run using the actual aspect solution of the \chandra\ observation, which sets the total number of X-ray counts per sky pixel, allowing each sky pixel to receive contributions from each lensed image.~The first ML analysis registers the four lensed images to the X-ray data, thereby correcting for the offset in \chandra\ absolute astrometry. We then construct a grid of putative X-ray source positions within the caustic. The second ML test repeats the first ML test at each putative X-ray source position. The putative position where the ML is minimal localizes the most likely X-ray source. Exploiting the theorem that the difference in ML is distributed as $\chi^2$ \citep{cash_1979_likelihood}, for two degrees of freedom (the RA, DEC of each alternate X-ray source position), we subsequently infer confidence contours for the X-ray source position at given confidence intervals~(Figure \ref{fig:results_fullband}).

\section{Mass model for J0659 from \gaia\ and \hst}
\label{sec:gaia} 

We summarize the findings of previous works on J0659. We then describe our model for the projected surface mass density of the system lensing J0659.~We refer to Sec. \ref{subsec:results_massmodel} for a discussion on the negligible impact of the degeneracies inherent to lensed quasar mass models for our astrometric analysis.

We perform lens modeling using the following cosmological parameter values: $H_0 = 67.5~\mathrm{km}~\mathrm{s}^{-1}~\mathrm{Mpc}^{-1}$, $\Omega_\mathrm{m} = 0.308$, $\Omega_\Lambda = 0.692$~\citep{planck_2016}.~To perform milliarcsecond X-ray astrometry relative to the optical, we require a lens model that reproduces the positions of \gaia\ DR3 quasar lensed image positions to milliarcsecond precision~(i.e.~beyond \gaia's astrometric precision).~Our mass model complies with this requirement, enabling source plane reconstruction, which provides the localizes the spatial origin of the emission yielding the \gaia\ DR3 data.

\subsection{Discovery and previous works on J0659} 
\label{subsec:orions_intro_gaia}
J0659 was discovered independently by \citet{delchambre_2019_j0659+disco} and \citet{lemon_2023_j0659+disco} after~\gaia\ Data Release 2~(\gaia\ DR2). \citet{delchambre_2019_j0659+disco} performed a blind search for lensed quasars in \gaia\ DR2 using a supervised machine learning algorithm trained on astrometric and photometric \gaia\ data. Although the \gaia\ DR2 data only shows three lensed images, the PanSTARRS image \citep[figure 1 of][]{stern_2021_j0659+massmod} reveals four lensed images forming a kite pattern \citep[for which J0659 was later referred to as `Orion's crossbow';][]{stern_2021_j0659+massmod}.~In parallel, \citet{lemon_2019_j0659+disco}~also conducted a search for lensed quasar candidates in \gaia\ DR2, subsequently confirming the quadruply lensed nature of J0659 in~\citet{lemon_2023_j0659+disco} with spectroscopic follow-up.

J0659 is classified as radio-quiet \citep{gordon_2020_radio, stern_2021_j0659+massmod,connor_2022_J0659+xmm}, but all four components were detected with the VLA at $6 \ \mathrm{GHz}$, with a total flux of $0.37 \ \mathrm{mJy}$ \citep{Dobie2024}. Their Figure 2 hints at multiple components for the A and D images.

The first mass models --models for the projected surface mass distribution of the deflector system-- of J0659 were presented by 
\citet{stern_2021_j0659+massmod} based on observed positions and fluxes of the lensed images in the Sloan Digital Sky Survey \citep[SDSS;][see their table 5 for the input data]{stern_2021_j0659+massmod}. The authors noted that the PanSTARRS image of J0659 shows four distinct blue components (each of the quasar lensed images) and a central image of the lensing galaxy \citep[see figure 1 of][]{stern_2021_j0659+massmod}.~Using \keck\ spectra obtained at different position angles, \citet{stern_2021_j0659+massmod}~spectroscopically confirmed that: the blue components correspond to a quasar at a source redshift of $z_\mathrm{s} = 3.083$, the red component corresponds to an early-type lensing galaxy at $z_\mathrm{L} = 0.766$, and that the quasar exhibits various absorption-line systems, although the lensing galaxy does not. \citet{stern_2021_j0659+massmod} noted that a mass model consisting of a single deflector modeled with either a Singular Isothermal Sphere (SIS) with external shear \citep[SIS + $\gamma$; see table 4 of][]{stern_2021_j0659+massmod} or a Singular Isothermal Ellipsoid (SIE) with external shear \citep[SIE + $\gamma$; see table 5 of][]{stern_2021_j0659+massmod} to capture the $z_\mathrm{L}=0.766$~early-type lens is unable to reproduce SDSS  quasar lensed image positions and fluxes,~suggesting the presence of a nearby isolated galaxy along the line-of-sight.

\begin{figure}[ht!]
\plotone{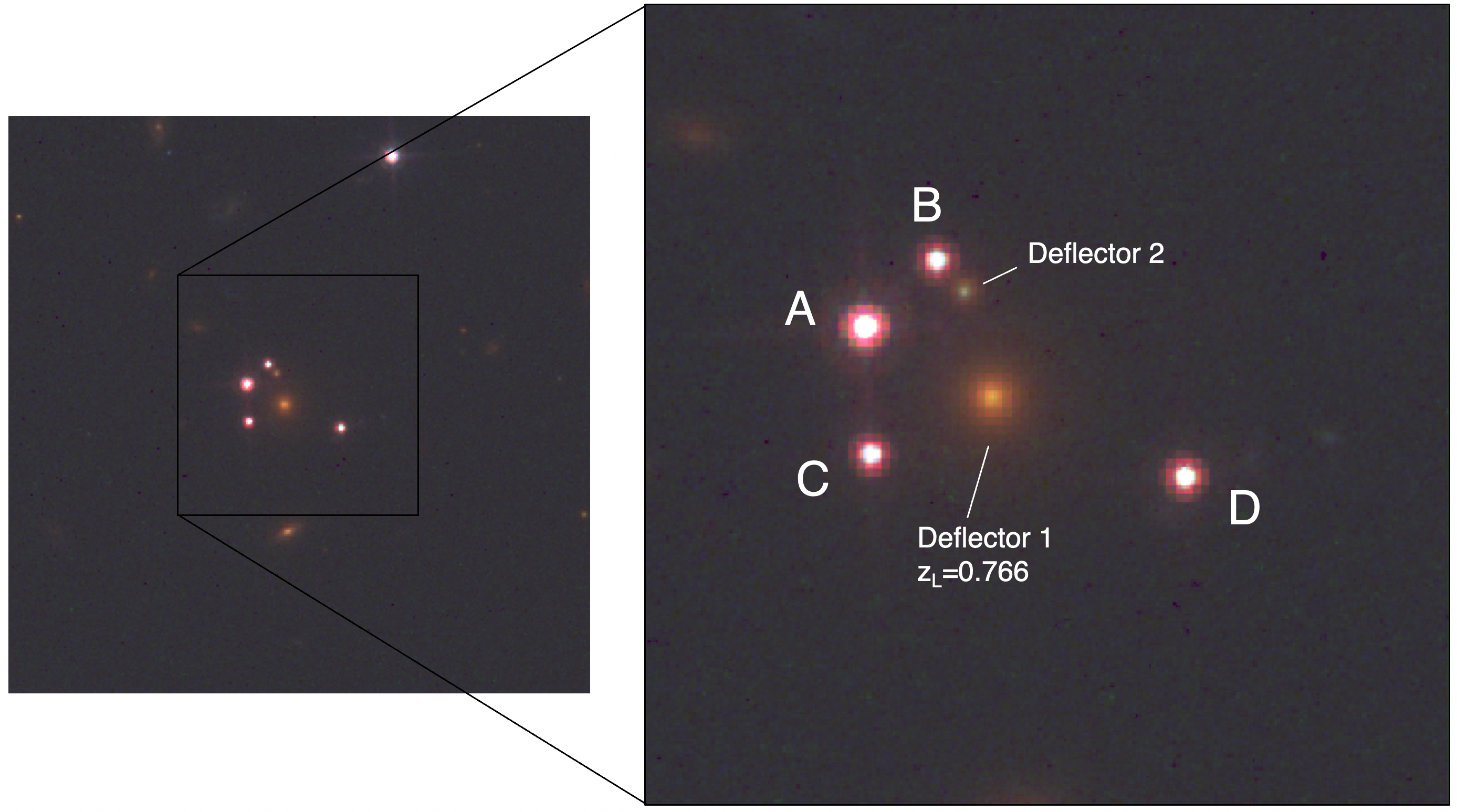}
\caption{Left:~red, green,~and blue (RGB) image of Orion's crossbow and its field using the \hst\ filters F160W (red), F814W (green) and F475W (blue).~The image is 0.5 arcmin $\times$ 0.5 arcmin.~Right:~Zoom on the system (0.2 arcmin $\times$ 0.2 arcmin). The labeling of the four quasar lensed images and of the two deflector systems is kept throughout the text.~Both images are centered at J2000 RA 06h 59m 03.971s and Dec +16$^{\circ}$ 29' 08.865", where North is up, East is left.}.
\label{fig:orions_optical_image}
\end{figure}

Figure \ref{fig:orions_optical_image} shows that the \hst\ image of J0659 reveals a deflector near `image B'.~Observations of J0659 taken across multiple filters by Wide-Field Camera 3 aboard \hst~were used by \citet{schmidt_2023_j0659+massmod} to model both the lens and source light and the lens mass distributions.~The authors employed a mass modeling pipeline applied to thirty-one quadruply lensed quasars observed by WFC3/\hst.~Table 1 of \citet{schmidt_2023_j0659+massmod} shows free parameters which were fitted using a $\mathrm{SIE} + \gamma$ parametrization \citet{schmidt_2023_j0659+massmod}, with associated statistical uncertainties $< 3\%$. For this reason, J0659 was subsequently identified as a compelling candidate for time delay cosmography investigations.~Gravitational time delays from J0659 from monitoring on the VLT Survey Telescope were recently reported by \citet{dux_2025_j659+tdels}.

\subsection{Model for the projected surface mass density of the lens from \gaia\ DR3 and \hst }

Guided by the observed positions of the four lensed images of J0659 reported by \gaia\ DR3 and informed by the presence of the additional deflector revealed by WFC3/\hst\ observations, we performed mass modeling in J0659 using the parametric lensing software \gravlens\ \citep{keeton-gravlens-2001,keeton-gravlens-2001-ii}. A parametric (non-pixelated) model reconstruction is expected to be suited for describing the lensing mass distribution since the quasar lensed images are point-like.~As required by the milliarcsecond X-ray astrometry we want to perform, we infer a mass model so that the eight position coordinates of the four lensed images observed by \gaia\ DR3 are reproduced to within one milliarcsecond.

We highlight that we have \emph{not} attempted to reproduce the observed image fluxes (or their corresponding flux ratios) since these may be affected by microlensing by stars in the lensing galaxy, substructure along the line-of-sight, and intrinsic source variability \citep{schechter_2002_microlensing,kochanek_2004_micro,kochanek_2004_subs,blackburne_2011_intrinsic,mosquera_2011_microlensing,nierenberg_2014_substructure}. Moreover, if microlensing were present, one may expect a shift in the observed image positions by up to $10^{-3} \times 0\farcs001$, mapping to the same astrometric shift in source plane reconstruction. This is three orders below the milliarcsecond precision of our mass model and is therefore negligible. We refer to \citet{vernardos_2024_microlensing-revi} for a recent review of how predictions on the effective convergence and shear maps of a swarm of microlensed images can be used to correct the strong lensing magnification affected by microlensing.

\begin{table*}
\centering
\begin{tabular}{|l|c|c|c|c|c||c|c|r|}
\hline
& Gaia Source ID & $\Delta \mathrm{RA}_\mathrm{A}[{\arcsec}]$ & $\Delta \mathrm{DEC}_\mathrm{A}[{\arcsec}]$ & $\delta x (\delta y)~[{\arcsec}/\mathrm{yr}]$ & $g$-mag & Predicted $\Delta\mathrm{RA}_\mathrm{A}[{\arcsec}]$ & Predicted $\Delta\mathrm{DEC}_\mathrm{A}[{\arcsec}]$\\
\hline

A & 3361094861567546112 & 0.0000 & 0.0000 & 0.0002 ($< 10^{-4}$) & 18.548035 & -0.0006 & 0.0005\\
B & 3361094865862486656 & 1.1115 & 0.9865 & NA~(NA) & 20.146595 & 1.1116 & 0.9864  \\
C & 3361094865862487168 & 0.0883 & -1.9012 & 0.0007 (0.0002) & 19.910759 & 0.0883 & -1.9012  \\
D & 3361094861571960704 & 4.9525 & -2.2354 &  0.0007 ($<10^{-4}$) & 19.966488 & 4.9525 & -2.2354 \\
\hline

\end{tabular}
\caption{Input \gaia\ DR3 data (\gaia\ source ID, RA, DEC, mean photometric $g$-band magnitude) and model predictions (RA, DEC) for the four quasar lensed images. The shifts in RA are calculated by subtracting the RA of each image from that of image A (Figure \ref{fig:source_image_plane}).~The shifts in DEC are calculated by subtracting the DEC of image A from the DEC of each lensed image.~The \gaia\ Archive lists the position of image A to be at RA = 104.767318 degrees and DEC = 16.486129 degrees.~The astrometric errors in RA, DEC (quoted as $\delta x, \delta y$, in units of milliarcsecond per year) were provided by \gaia\ DR3 for each image except for image B, where these data are not available (NA). These were conservatively assigned those of image D throughout the mass modeling (the third brightest image).}
\label{tab:gaia_observables_predictions}

\end{table*}

Table \ref{tab:gaia_observables_predictions} shows the input data we used to conduct our mass modeling for J0659. The coordinates of each of the four \gaia\ DR3 lensed quasar images are quoted with reference to those of the brightest image (A in Figure \ref{fig:orions_optical_image}).

Our mass model consists of two point-like deflectors lensing the background quasar \citep[at a source redshift $z_\mathrm{s}=3.083$;][]{stern_2021_j0659+massmod}. The first deflector (`deflector 1') corresponds to the early-type lensing galaxy at $z_\mathrm{L}=0.766$ and is parametrized by a~$\mathrm{SIE} + \gamma$ model.~The second deflector (`deflector 2') captures the point-like deflector close to image B and is parametrized as an SIS.~The best-fit lens model was found by solving the lens equation in \gravlens\ in both the source and image planes, following the guidelines in \citet{keeton-gravlens-2001}. Table \ref{tab:massmodel_parameters} shows the best-fit parameter values of our mass model, yielding a global chi-square value of $\sim 10^{-3}$.~We note the best-fit values for the early-type galaxy are broadly consistent with those reported by \citet{stern_2021_j0659+massmod}.~Section \ref{subsec:results_massmodel} further discusses our mass model.

\begin{table*}[t]
\center 
\centering
\begin{tabular}{|l|c|c|c|c|c|c|c|c|r|}
\hline
Component & $z$ & $\theta_\mathrm{E}$ & $\epsilon$ & $\mathrm{PA}_\epsilon[\mathrm{deg}]$ & $\gamma$ & $\mathrm{PA}_\gamma[\mathrm{deg}]$ & $\Delta \mathrm{RA}_\mathrm{A} [\arcsec]$ & $\Delta \mathrm{DEC}_\mathrm{A}[\arcsec]$ \\
\hline
Deflector 1 & 0.766 & 2.420 & 0.251 & -9.285 & 0.066 & -40.02 & 1.9838 & -1.0324 \\
Deflector 2 & NA & 0.010 & 0 & - & - & - & 1.5626 & 0.8859 \\
Source & 3.083 & - & - & - & - & - & 2.4222 & -1.1742\\
\hline
\end{tabular}
\caption{Best-fit values for the $\mathrm{SIE}+\gamma$~(early-type lensing galaxy)~and $\mathrm{SIS}$~(line-of-sight deflector close to image B)~in our mass model for J0659.~We include redshifts for the main deflector and the source \citep{stern_2021_j0659+massmod}, although these values were not employed during model optimization. The redshift of `deflector 2' close to image B is currently unknown from literature studies (NA).~The source redshift was fixed at $z_\mathrm{s}=3.083$ \citep{stern_2021_j0659+massmod}. The two deflectors follow the naming convention adopted in Figure~\ref{fig:orions_optical_image}. Coordinate shifts in both RA and DEC are quoted with respect to those of the brightest image, A, following the same sign convention applied in Table \ref{tab:gaia_observables_predictions}.}
\label{tab:massmodel_parameters}
\end{table*}

\begin{figure*}[ht!]
\plotone{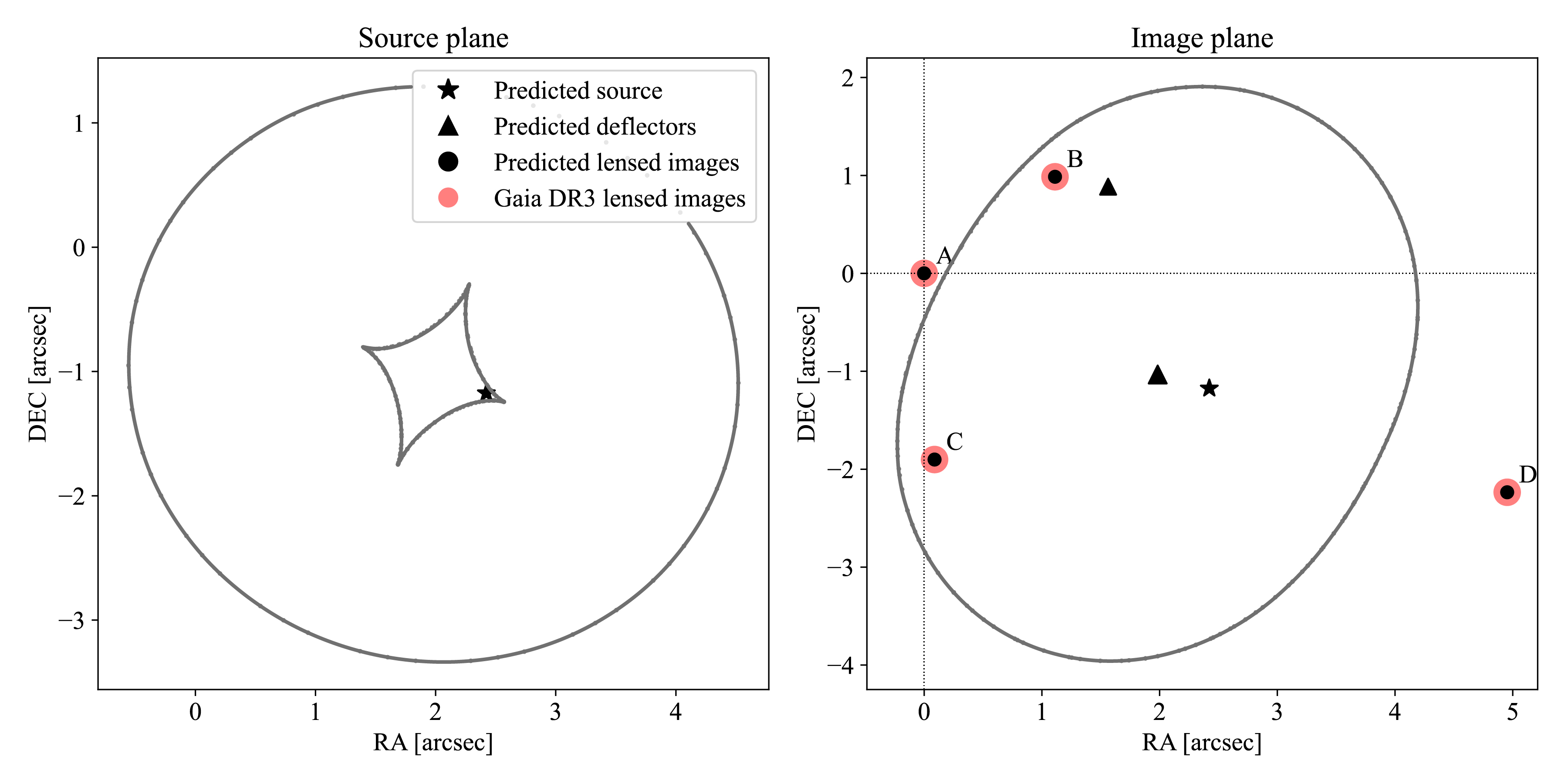}
\caption{Source plane (left) and image plane (right) reconstructions of the lensed system J0659. The source position predicted by our mass model is tens of milliarcseconds from the inner caustic, making J0659 an ideal candidate for the caustic method \citep{barnacka_2018_rev}. The image plane reconstruction shows that the \gaia\ DR3 observed lensed images and the model-predicted lensed images agree to better than one milliarcsecond. The model-inferred positions for the two deflectors in the lens model are shown. Coordinates are chosen so that the brightest image, A, is at (0,0), using the signs convention shown in Tables \ref{tab:gaia_observables_predictions} and \ref{tab:massmodel_parameters}.}
\label{fig:source_image_plane}
\end{figure*}

Table \ref{tab:gaia_observables_predictions} shows that our mass model for the lensing mass distribution in Orion's crossbow reproduces the observed positions of the four lensed images to milliarcsecond precision -- the first to use \gaia\ DR3 data. Figure \ref{fig:source_image_plane} shows the source plane and image plane reconstructions from our mass model, showing the lensing caustics and critical curves. The model predicts the source to be at a shortest distance of~$\sim 0\farcs051$~from the inner caustic~(see also Figure \ref{fig:results_fullband}). 

\section{Locating the origin of the X-ray emission in J0659} \label{sec:chandra}

Section \ref{subsec:x-ray_data_analysis} summarizes our spectral analysis of the archival \chandra\ observations of Orion’s crossbow.~Section~\ref{sec:x-ray_raytrace_sims}~introduces the ray-trace simulations of these data we performed to predict what each of the four X-ray lensed images seen by \chandra\ would look like if detected at ICRS \gaia\ DR3 lensed image positions. 

J0659 was observed by the Advanced CCD Spectrometer (ACIS-S) aboard \chandra\ over 15 - 16 January 2021. The source was observed in two separate segments~(ObsIDs: 22018 / 23825; exposure times: 1.67 / 14.87 ks; PI: D. Pooley),~accessible via\dataset[doi:10.25574/cdc.45]{https://doi.org/10.25574/cdc.459}.~In the remaining sections, we specify the results inferred with the longest of these observations, since, given its low signal-to-noise ratio (S/N), the inclusion of ObsID 22018 does not change our astrometric results.

\subsection{Spectral analysis of the archival \chandra\ data}
\label{subsec:x-ray_data_analysis}

\begin{figure}[ht!]
    \centering

    \newlength{\commonheight}
    \setlength{\commonheight}{6cm}

    \newlength{\commonwidth}
    \setlength{\commonwidth}{8cm}

    \includegraphics[height=7.2cm,width=11cm]{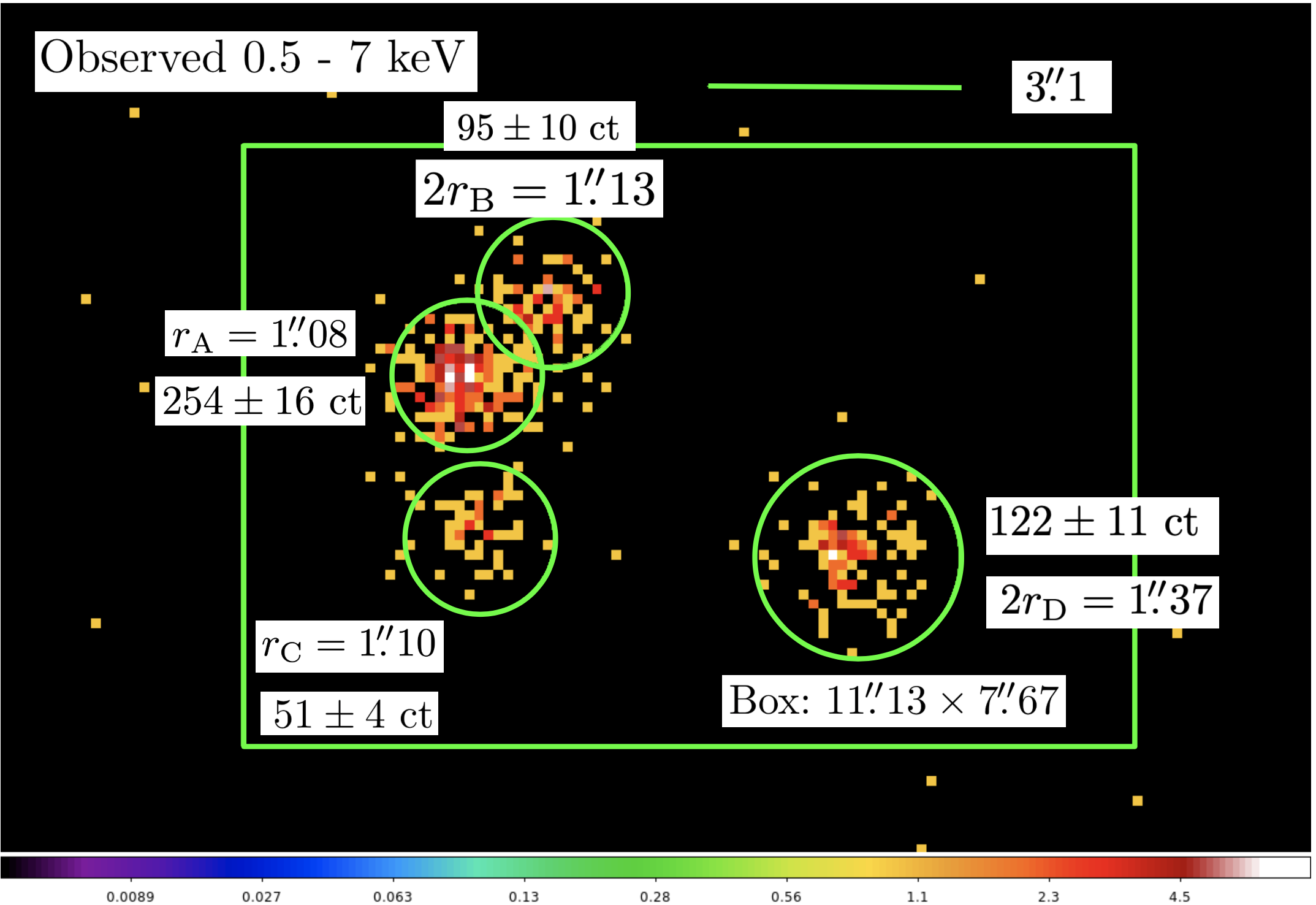}

    \vspace{0.5em} 
    
    \includegraphics[height=6.5cm,width=11cm]{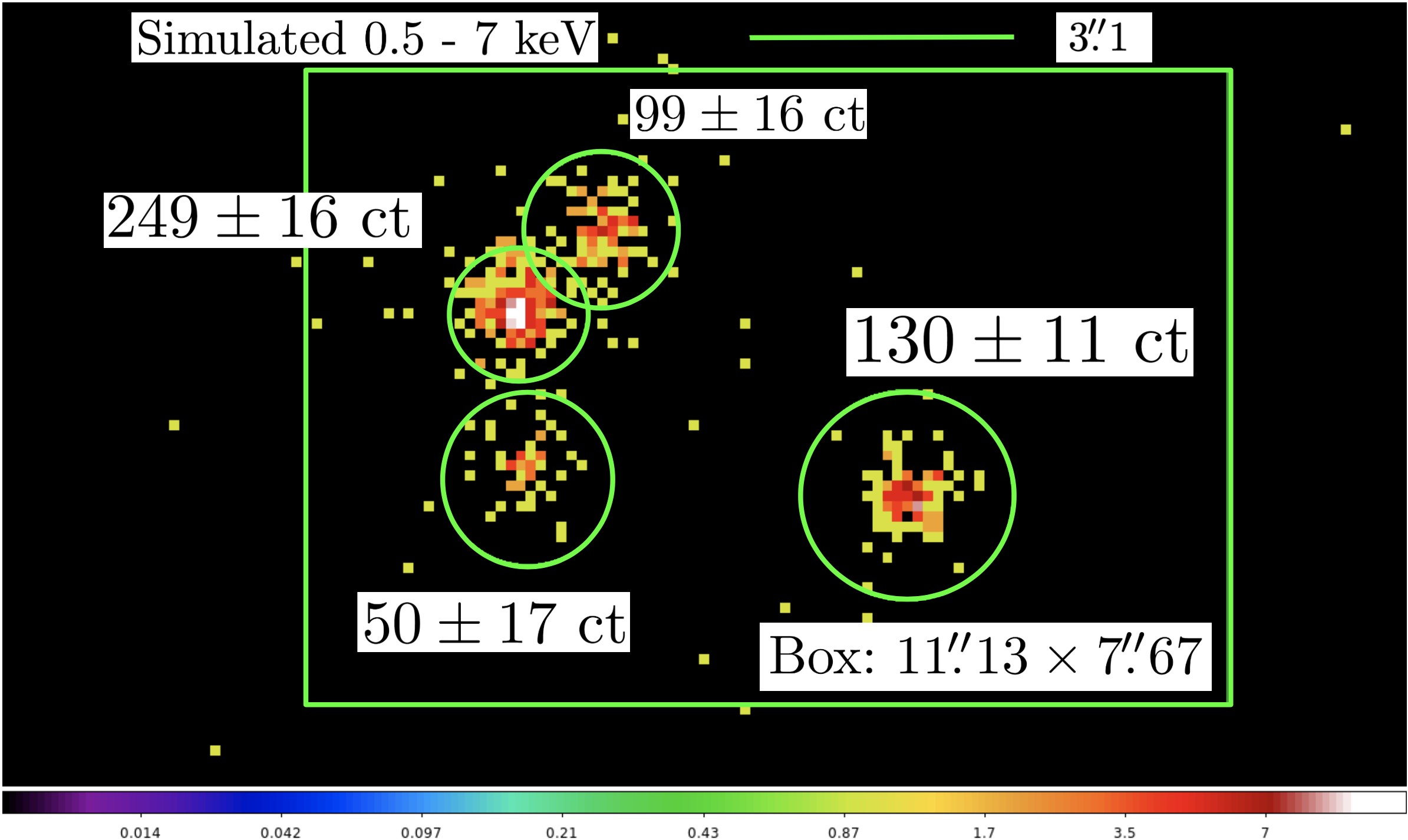}

    \caption{Top: 0.5–7 keV \chandra\ image of the quadruply lensed quasar J0659 (ObsID: 23825), binned in $0\farcs246$ (half of the size of a native ACIS pixel). The radius of each extraction region ($i$-th image) is indicated by $r_i$.~Error bars in the number of counts correspond to $1\sigma$ statistical uncertainties on the number of photon counts enclosed by each region.~The box encloses $516 \pm 23$ X-ray photons. The color scale is logarithmic. Bottom: Composite 0.5–7 keV image showing one of the 1\,000 simulations run for each lensed image. The box enclosing the four lensed images in each frame is the same size as in the top.}
    \label{fig:chandra_data_simulations}
\end{figure}

We reprocessed the archival \chandra\ observations of J0659 using \ciao’s v14.6~\texttt{chandra\_repro} script.~Figure \ref{fig:chandra_data_simulations} shows the 0.5 - 7 keV~(observed frame)~image for ObsID 23825, showing four distinct emission regions.~Each of these regions is shown by the upper panel of Figure~\ref{fig:chandra_data_simulations}.~The centroids of images A and D are separated by $5\farcs2$, coinciding with the maximum image separation in the quadruplet observed by PanSTARRS \citep{stern_2021_j0659+massmod} and WFC3/\hst~(Fig.~\ref{fig:orions_optical_image}).~Each of the four X-ray lensed images is detected at high significance above the background level of $1.46 \times 10^{-3} \ \mathrm{counts}~\mathrm{arcsec}^{-2}$.~Figure \ref{fig:chandra_data_simulations} shows the box enclosing the observed X-ray lensed images used in our double maximum likelihood approach \citep[for a detailed explanation, see the Appendix of][]{GLAD+rogers_2025_he0435}.

We extracted the X-ray spectrum of each of the four circular regions in Figure \ref{fig:chandra_data_simulations} using \ciao’s \texttt{specextract} tool.~While there is some overlap between the photon counts enclosed by images A and B, we note this does not affect our results, since our double maximum likelihood approach~(outlined in Section \ref{sec:methodology})~computes the number of counts per bin, comparing with the number of counts per pixel bin in the observed data by considering a normalization parameter for each putative source position.~We used \xspec\ to group each of the four spectra to ensure a minimum of one photon count per spectral bin. We then used \sherpa\ to fit each grouped spectrum, finding that all spectra are well described by a primary continuum power-law modified by the effects of Galactic absorption along the quasar line-of-sight.

Table \ref{tab:bestfit_params_spectra_chandra} lists the 68.3\% confidence intervals of the spectral index of the power-law component (note its normalization component was also fitted for).~We modeled the effects of Galactic absorption using the \texttt{tbabs} multiplicative model of \citet{wilms_2000_tbabs}, assuming the element abundance ratios of \citet{anders_grevesse_1989_nh+abund} and assuming a fixed neutral hydrogen column density to $N_\mathrm{H} = 1.16\times 10^{21} \ {\mathrm{cm}}^{-2}$ \citep[][]{hi4pyCollab_2016_nh+ref}.~The model-predicted (magnified) $0.5 - 7 \ \mathrm{keV}$ broadband fluxes are listed in Table \ref{tab:bestfit_params_spectra_chandra}. 

We note that \citet{connor_2022_J0659+xmm} undertook a joint spectral analysis of the four lensed images from a 10 ks \xmmnew\ observation.~The authors reported an absorption-corrected magnified 2-10 rest-frame luminosity of $\mathrm{log}_{10}(L_{2-10}[\mathrm{erg~s^{-1}}])= 46.44 \pm 0.02$ from the four joint lensed images (unresolved by \xmm).~This value is consistent with the 2-10 keV rest-frame luminosity the fluxes in Table \ref{tab:bestfit_params_spectra_chandra} correspond to when added together.

Section \ref{sec:x-ray_raytrace_sims} describes how the (energy-dependent) model-predicted flux density of the spectral model of each lensed image was used to perform the ray-trace simulations. We note that these model-predicted fluxes are consistent with those featured in the \chandra\ Source Catalog v2.1~\citep[CSC2.1;~see][]{evans_2024_csc}, as follows.~The CSC2.1 lists three individual source detections\footnote[1]{See \href{doi:10.25574/csc2}{doi:10.25574/csc2} \citep{csc_2} to access the second release of the \chandra\ Source Catalog for information on the criteria on which individual sources are detected.} in the direction of J0659. These three sources correspond to three out of the eight lensed images of the quasar detected with both ObsIDs. For each of the three images in the Catalog, the lower and upper fluxes are consistent with the model-predicted fluxes inferred from our spectral models as listed in Table \ref{tab:bestfit_params_spectra_chandra}. In addition, we note we found no evidence for intrinsic quasar variability from comparing the X-ray fluxes corresponding to each lensed image in the two archival \chandra\ observations of J0659 (i.e.~Obs ID 22018 and 23825).

\begin{table*}
\centering
\begin{tabular}{|l|c|c|r}
\hline
Image & $\Gamma_X$ & $F_\mathrm{0.5 - 7}~[\mathrm{erg}~\mathrm{s}^{-1}~\mathrm{cm}^{-2}]$ 
\\
\hline
A & $1.53 \pm 0.13$ & $(2.69\pm 0.43) \times 10^{-13}$ \\
B & $1.64\pm 0.24$ & $(9.35\pm 0.97) \times 10^{-14}$ \\
C & $1.67\pm 0.31$ & $(5.41\pm 0.10)\times10^{-14}$ \\
D & $2.12\pm 0.21$ & $(1.16\pm 0.13)\times 10^{-13}$\\
\hline

\end{tabular}

\caption{68.3\% confidence intervals for the power-law indices and broadband model-predicted fluxes from our analysis of \chandra\ ObsID 23825 for images A, B, C, and D. The statistical uncertainties on the spectral index were calculated with \sherpa,~whereas the uncertainty in the model-predicted fluxes were set by the square root of the number of counts in each lensed image (see Figure \ref{fig:chandra_data_simulations}). Each image spectrum was fitted with a power-law continuum~(defined by the spectral index $\Gamma_X$ and a normalization parameter)~modified by the effects of Galactic absorption.}
\label{tab:bestfit_params_spectra_chandra}
\end{table*}

\subsection{Ray-trace simulations of the archival \chandra\ data}
\label{sec:x-ray_raytrace_sims}
We used the state-of-the-art \saotrace\ simulation package to perform the ray-trace simulations of the archival \chandra\ observations. This publicly available tool was built as a dedicated simulation package for the development, calibration, and analysis of the \chandra\ X-ray Observatory telescope.~We refer to section 3 of \citet{saotrace-jerius-2004} for a detailed description of how pre-flight and in-flight calibration measurements are incorporated into \saotrace\ and were validated for \chandra.~Although we carried out \saotrace\ simulations for both archival \chandra\ ObsIDs, here we present results for ObsID 23825 since this ObsID is the only one that impacts our astrometric results.

We carried out 1\,000 simulations of each lensed image for ObsID 23825.~For a given lensed image, \saotrace\ requires the following inputs to carry out a single simulation: 

\begin{itemize}
    \item The coordinates (RA, DEC) and roll angle of the nominal pointing direction of the actual ObsID.
    \item The actual aspect solution of the observation.
    \item The ICRS coordinates (RA, DEC) of the image in \gaia\ DR3.
    \item A prediction for the energy-dependent X-ray flux density of the given lensed image, which we assigned to the X-ray model-predicted flux density described in Section \ref{subsec:x-ray_data_analysis}.
    
\end{itemize} We performed 1\,000 ray-trace simulations of each lensed image in \chandra\ mirror coordinates. Each simulation incorporates the photon and particle background of the observation and uses its exact start time and duration and dither pattern but randomly changes the energy and direction of each X-ray incident into the telescope.~We subsequently converted each of these simulations into detector units (i.e.,~a level 2 events file in \emph{fits}~format) using the Energy-Dependent Subpixel Event Repositioning (EDSER) algorithm in \marx~\citep{li_2004_edser}. We then merged each of the 1\,000~\emph{fits}~files for each lensed image using \ciao’s \texttt{dmmerge} script. The merging process led to a single simulated image for each of the four lensed images. We expect the centroid error of each simulated image to be only 3\% of the error of the centroid of its corresponding actual image \citep{auchettl_2015_chandra+astro}.

Our methodology (described in Section \ref{sec:methodology})~involves forward-modeling our lens model (Section \ref{sec:gaia}) across a grid of putative source positions, exploiting caustics as nonlinear amplifiers, since each putative source position will predict a distinct set of observables (positions and magnifications of the four lensed images). Changes in image magnification will be more noticeable for putative source positions in the direction perpendicular to the caustic \citep[][]{barnacka_2018_rev}.

We constructed a grid of 200 trial source positions (`putative sources')~along and perpendicular to the caustic. The position of the optical source predicted by our mass model~(i.e.~the `optical source';~Table \ref{tab:massmodel_parameters})~is shown by the black star in Figure \ref{fig:results_fullband}.~Our grid of putative sources comprises an $\sim 0\farcs12 \times 0\farcs25$ area within the inner caustic (Fig. \ref{fig:results_fullband}).~The positions of the four lensed images predicted by forward-modeling our mass model are shown by the left panel of Figure \ref{fig:results_fullband},~which we found using~\gravlens’~\texttt{findimg3} command. 

\section{Results}  \label{sec:results}

Our methodology, summarized in Section \ref{sec:introduction}, registers the position of the quasar lensed images in the \chandra\ X-ray observations of J0659 to ICRS positions reported by~\gaia\ DR3.~Our mass model predicts the quasar lensed images to milliarcsecond precision of the actual \gaia\ DR3 observations. The combination of these two factors allows locating the origin of the broadband (0.5 - 7 keV)~X-ray emission in J0659 observed by \chandra~to milliarcsecond precision relative to the optical source.

  \begin{figure*}[ht!]
\plotone{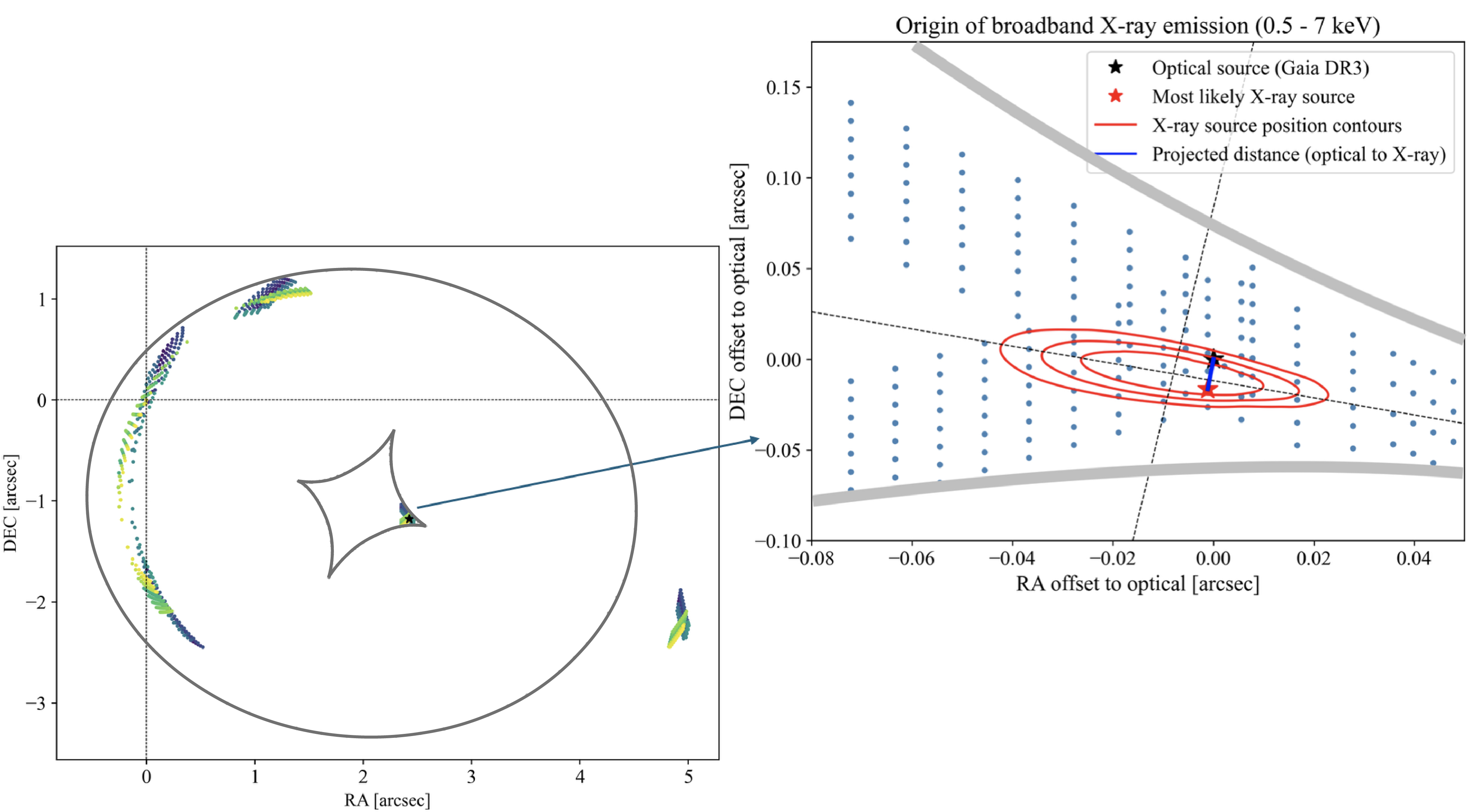}
\caption{Left:~Schematic showing the grid of putative sources we considered. Each putative source position is color coded according to its corresponding (forward-modeled) predicted lensed images. RA, DEC are the difference coordinates, referenced so that the position of the optical image A is at (0,0) and with RA increasing to the West. Right: zoom into the grid of putative source positions (in blue), defining a triangular region within the edge of the inner caustic. The 68.3\%, 95.5\% and 99.7\% CLs for the relative X-ray source position are shown by the red contours. The most likely X-ray source position (red star) is the putative position with the lowest value of the maximum likelihood. RA, DEC coordinates are quoted so that the predicted image position of the optical source is at (0,0),~as indicated by the dashed vertical and horizontal lines.~The projected distance between the most likely X-ray source and the optical, at $\approx 2\sigma$ confidence, is shown by the blue line.}
\label{fig:results_fullband}
\end{figure*}

Figure \ref{fig:results_fullband} shows the 68.3\%, 95.5\% and 99.7\% confidence level (CL) contours for the spatial origin of the broadband X-ray emission in the gravitationally lensed quasar J0659. The contours are derived from the chi-squared distribution obtained by subtracting the maximum likelihood value of the most likely X-ray position from the maximum likelihood assigned to each putative source position, for two degrees of freedom (the RA and DEC of the putative source).~Section \ref{sec:methodology} describes the purpose of assigning a maximum likelihood value to each putative source position.~Out of the 200 trial source positions, the putative source position with the lowest maximum likelihood value is the most likely position for the observed broadband \chandra\ X-ray emission (0.5 - 7 keV;~the `X-ray source' in Figure \ref{fig:results_fullband}).

Assuming the bulk of the X-ray emission in J0659 is attributable to a single position, Figure \ref{fig:results_fullband} locates the X-ray-emitting region in J0659 to within a ellipse with semi-major and semi-major axes $a \times b = $ \ellipsemajor\ $\times$ \ellipseminor~centered $0\farcs014$ away from the optical source at 68.5\% confidence. This corresponds to an area of $628 \ {\mathrm{milliarcseconds}}^2$. Figure \ref{fig:results_fullband} also shows the $2\sigma$ and $3\sigma$ confidence levels for the X-ray source position. The most likely source position does not coincide with the center of the ellipse. This must by coincidence, and is attributed to the nonlinearity of the caustic method, wherein changes in maximum likelihood are most significant in directions perpendicular to the caustic.

Figure \ref{fig:results_fullband} shows that the X-ray emission in J0659 is spatially consistent with the optical source at $\approx 2\sigma$ confidence. At the source redshift of $z_\mathrm{s}=3.083$,~the angular scale is $0\farcs133/\mathrm{kpc}$.~Therefore, the angular offset between the optical source and the most likely X-ray source position of $0\farcs016$ sets an upper limit of \offset\ between these two distinct emission regions in J0659 at the source redshift ($z_\mathrm{s} = 3.083$). Given the statistical power of the \chandra\ X-ray observations, the existence of an optical-X-ray offset in J0659 \textit{below} this projected distance cannot be currently ruled out. We discuss the significance of this upper limit in Section \ref{subsec:results_fullband}.

We have also performed, for the first time, milliarcsecond spectrally resolved X-ray astrometry and have demonstrated the applicability of our method to localize the soft and hard X-ray emitting regions in J0659. For these individual analyses, we applied the same methodology as for the broadband X-ray astrometric analysis but~\emph{only} selected the soft / hard photon events in the observed and simulated \chandra\ observations in the parts of the methodology described in Section \ref{sec:x-ray_raytrace_sims}~(observed energies: $0.5 - 2 \ \mathrm{keV}$ / $2 - 7 \ \mathrm{keV}$).

\begin{figure*}[ht!]
  \centering
  \includegraphics[width=0.6\textwidth,height=0.5\textwidth]{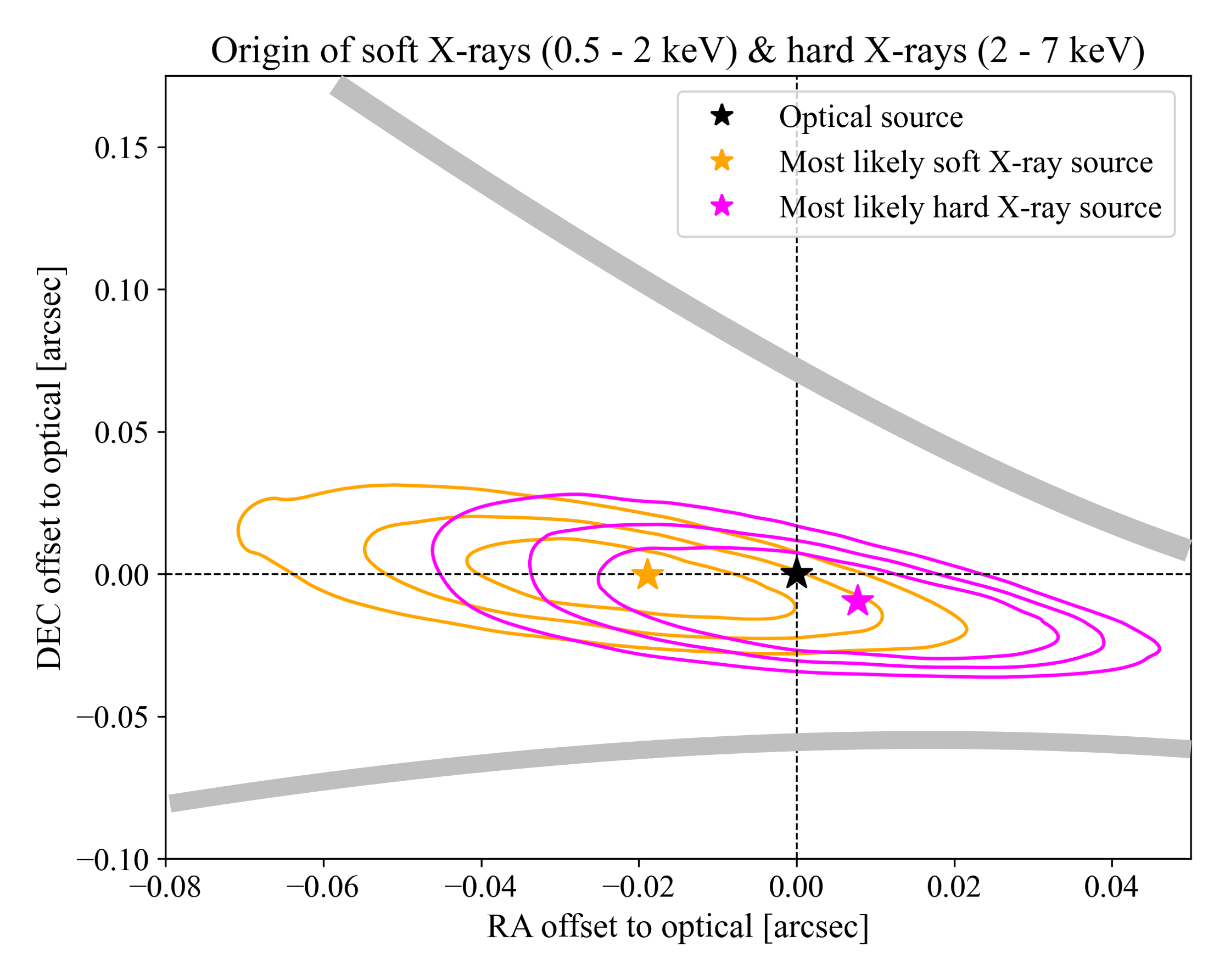}

  \caption{68.3\%, 95.5\% and 99.7\% CL contours for the origin of the soft X-ray emission (magenta) and the hard X-ray emission (orange) in J0659. The orange and magenta stars indicate the most likely origin of the soft and hard X-ray emissions in J0659, respectively. RA, DEC coordinates are quoted so that the position of image A at the optical source (represented by the black star)~is at (0,0). Lines connecting each quadrant of the inner caustic are over-plotted in gray.
  }
  \label{fig:results_soft-hard-bands}
\end{figure*}

Figure \ref{fig:results_soft-hard-bands} shows the contours corresponding to the 68.3\%, 95.5\% and 99.7\% CLs for the origins of the soft X-ray (0.5 - 2 keV) and hard X-ray (2 - 7 keV) emission in J0659 observed by \chandra.~As in the broadband analysis, these contours were inferred from their corresponding chi-square distributions.~In each case,~the maximum likelihood tests compare the number of counts per bin observed in the soft/hard bands to the number of counts per bin predicted at each putative source in the soft/hard bands.

As expected, both the soft X-ray and hard X-ray emitting regions in J0659 are consistent with the origin of the broadband X-ray emission shown by Figure \ref{fig:results_fullband}.~Figure \ref{fig:results_soft-hard-bands} locates the distinct soft and hard X-ray-emitting regions in J0659 to within a \ellipsemajor$\times$\ellipseminor~ellipse.~Both emission regions are consistent with the optical source at $95.5\%$ confidence.

\section{Discussion}   \label{sec:discussion}

 \subsection{Lens mass model for J0659 from \gaia/DR3 and \hst}
  \label{subsec:results_massmodel}
Section \ref{sec:gaia} introduced our model for the mass distribution lensing J0659. Although the quasar was discovered independently by \citet{delchambre_2019_j0659+disco} and \citet{lemon_2023_j0659+disco}~from \gaia\ DR2 observations, ours is the first mass model derived from \gaia\ DR3 and \hst\ observations. Our model reproduces the observed positions of the four \gaia\ DR3 lensed images to milliarcsecond precision. %

Table \ref{tab:massmodel_parameters} shows the best-fit values of the free parameters of our mass model. This consists of an SIE + $\gamma$ component and an independent SIS component. The $\mathrm{SIE \ +} \ \gamma$ component captures the early-type elliptical galaxy at $z_\mathrm{L} = 0.766$ (`deflector 1' in Figure \ref{fig:orions_optical_image})~which was spectroscopically confirmed by \citet[][]{stern_2021_j0659+massmod}. An SIE parametrization is widely used for early-type deflectors and is expected to capture baryonic effects at the centers of elliptical galaxies \citep[see][and references therein]{lemon_2024_review+strongLensing}. The external shear is motivated by the slight asymmetries in the lensing configuration of J0659 \citep{stern_2021_j0659+massmod}.~The SIS component captures the additional deflector along the line-of-sight revealed by the WFC3/\hst\ observations \citep[`deflector 2' in Figure \ref{fig:orions_optical_image};~see][]{schmidt_2023_j0659+massmod}. The main deflector~($z_\mathrm{L}=0.766$)~has an Einstein radius of $\theta_\mathrm{E}=2\farcs42$,~corresponding to a total enclosed mass of $\sim 10^{13} M_\odot$, as expected for a galaxy with a massive, old stellar population. In addition, if located at $z=0.766$, the Einstein radius of the secondary deflector ($\theta_\mathrm{E}=0\farcs01$) would correspond to a total enclosed mass of $1.2\times 10^{10} M_\odot$, which would suggest this additional deflector is an LMC-type satellite galaxy associated with the main deflector. 

We emphasize that the impact of the adopted lens mass model on our astrometric analysis is negligible. This was demonstrated by \citet{GLAD+rogers_2025_he0435} for the quadruply lensed quasar HE 0435 (Tab. \ref{tab:offset_quads}). The authors showed that different physically-motivated models that reproduce the observed \gaia\ DR3 image positions of HE 0435 yield only minor differences in image positions and magnifications, corresponding to offsets a few tenths of a milliarcsecond in the source plane.~These offsets are negligible compared to the X‑ray astrometric uncertainties in the image plane. Furthermore, since strong lensing is achromatic, any lens model that matches the \gaia\ DR3 positions will also predict the \chandra\ lensed image positions to the same astrometric precision, provided the \gaia\ and \chandra\ sources are spatially coincident. Therefore, the dominant limitation to achieving milliarcsecond X‑ray astrometry in J0659 is the signal‑to‑noise of the \chandra\  observations, rather than the choice of lens model, as long as the latter is motivated by the observations and reproduces the lensed image positions within a milliarcsecond of the \gaia\ DR3 data. While galaxy–galaxy gravitational lensing models are subject to intrinsic degeneracies \citep[including the mass-sheet degeneracy; ][]{schneider_2013_mass-sheet} and our mass model may not be unique, it critically reproduces the observed \gaia\ DR3 lensed image positions within a milliarcsecond and is consistent with other mass model in the literature \citep[][]{stern_2021_j0659+massmod,schmidt_2023_j0659+massmod}. Specifically, the free parameters of our model lie within the error bars of the mass model parameters in the literature \citep{schmidt_2023_j0659+massmod}.

\begin{table*}
\centering
\begin{tabular}{|l|c|c|c|c|c|r|}
\hline
Source & Source redshift, $z_\mathrm{s}$ & Offset [arcseconds] (projected distance) & Reference\\ \hline 
J~2019+1127 & 3.273 & $0\farcs022$ (175 pc) & \citet{GLAD+schwartz_2021_j2019};~Radio vs.~X-ray \\

 CLASS B0712+472 & 1.340 & $0\farcs011$~(95 pc) & \citet{GLAD+spingola_2022_2systems};~optical vs.~radio\\
 CLASS B1608+656 & 1.394 & $< 0\farcs009$ (78 pc) & \citet{GLAD+spingola_2022_2systems};~optical vs.~radio\\
HE 0435-1223 & 1.689 & $0\farcs003$ (150 pc) & \citet{GLAD+rogers_2025_he0435};~optical vs.~X-ray\\
GraL J~0659+1609 & 3.083 & $<0\farcs016$ (\offset) & This work;~optical vs.~X-ray \\
\hline
\end{tabular}
\caption{Sample of quadruply lensed quasars whose offsets between the optical and/or radio and X-ray emitting regions we have determined to date.~The approximate projected separation between the two emission regions of interest (determined at either the $1\sigma$, $2\sigma$, or $3\sigma$ confidence levels) vs.~source redshift for each lensed quasar is shown by Figure \ref{fig:chen_updated}.~For the dual AGN candidate J2019, we quote the projected separation of the two AGN cores reported by \citet{spingola_2019_j2019_vlbi}.}
\label{tab:offset_quads}
\end{table*}

 \subsection{Broadband and spectrally resolved X-ray astrometry of J0659 at milliarcsecond resolution}
 \label{subsec:results_fullband}

We have employed archival \chandra\ X-ray observations of J0659 to perform milliarcsecond X-ray astrometry of this lensed quasar.~Our methodology leverages \gaia\ astrometry, \chandra\ imaging, and caustics as nonlinear amplifiers,~allowing us to improve \chandra's spatial resolution at cosmological distances by up to two orders of magnitude, contingent on the strong lensing magnification \citep{barnacka_2018_rev}.~The boost in spatial resolution provided by our methodology scales linearly with both the square root of the number of photon counts in the \chandra\ data and the square root of the strong lensing magnification $\mu$. Using the $\mu\sim 32$ prediction of our mass model and the source redshift $z_\mathrm{s} = 3.083$,~our approach therefore boosts \chandra's spatial resolution by a factor following the square root of the number of photons in the \chandra\ observations, thus by a factor $\sim 100$. 

Figure \ref{fig:results_fullband} shows that, if attributable to a single source position, the X-ray emitting region in J0659 is consistent with the optical source to the 99.7\% level. In addition, the X-ray source is located within a $0\farcs020\times 0\farcs010$ ellipse at $1\sigma$ confidence.~The most likely X-ray source is spatially consistent with the optical at $\approx$ the 95.5\% level. At this level of confidence, the maximum possible distance between the X-ray and optical emission regions at $z_\mathrm{s} = 3.083$ is \offset. If there indeed exists an X-ray-to-optical spatial offset, it would have to be at a projected distance \emph{below}~\offset\ as the caustic method is unable to resolve such an offset given the statistical power of the archival \chandra\ data.

Table \ref{tab:offset_quads} lists X-ray/radio to optical offsets determined with the caustic method on five quadruply lensed quasars with archival \chandra\ and \hst\ or \gaia\ observations.~These quasars span a redshift range $z_\mathrm{s} \sim 1.30 - 3.38$, including J0659, which is one of the highest-$z$~systems whose structure has yet been explored.

For the first time, we have demonstrated that our methodology can pinpoint the distinct spatial origins of soft and hard X-ray emission in quadruply lensed quasars to milliarcsecond precision.~Figure \ref{fig:results_soft-hard-bands} shows that both the soft and hard X-ray emitting regions in J0659 are spatially coincident with the optical source to the 99.7\% level. Nevertheless, the ellipses defining the confidence contours for both separate emissions do not fully overlap and are centered at different locations relative to the optical source. We highlight that there is a $0\farcs020 \times 0\farcs010$ region where the soft X-ray component could originate which is excluded by the hard X-ray emission region to 99.7\% confidence. 

\begin{figure}[ht!]
    \centering
    \setlength{\commonheight}{6cm}

    \includegraphics[height=\commonheight,keepaspectratio]{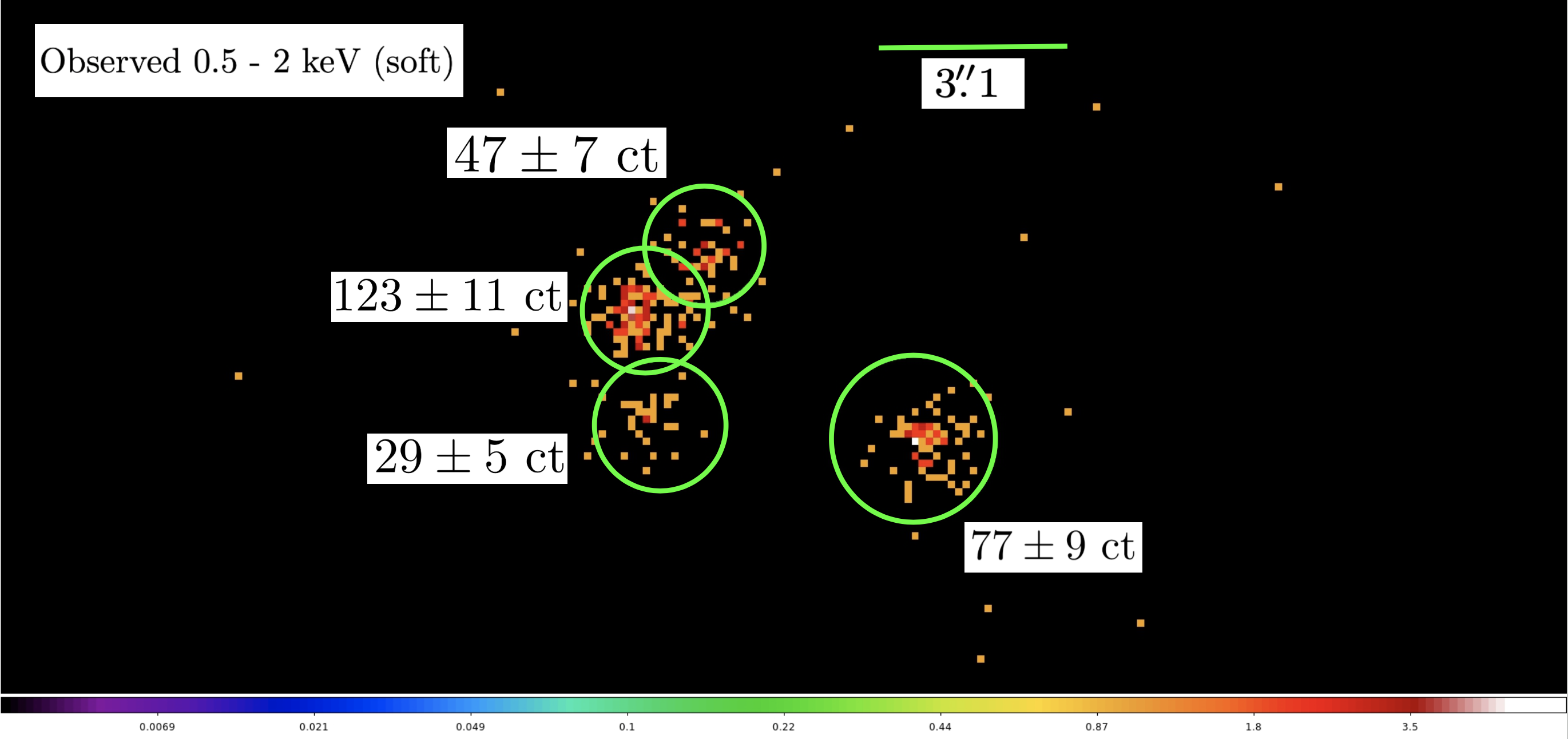}

    \vspace{0.5em} 
    
    \includegraphics[height=\commonheight,keepaspectratio]{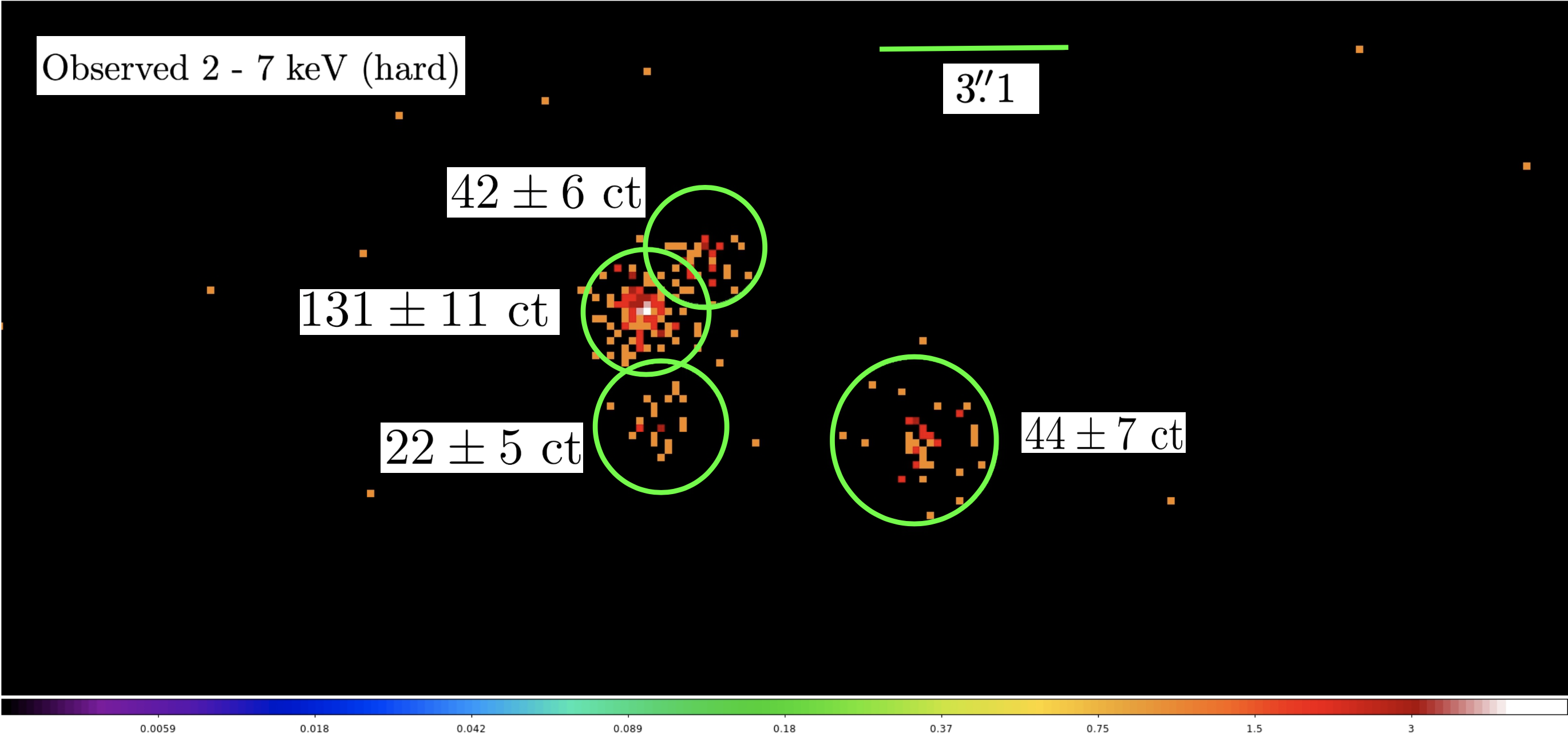}

    \caption{\chandra\ image of the quadruply lensed quasar J0659 (ObsID: 23825), binned in $0\farcs25$ (i.e., half the size of a native ACIS pixel), for soft (0.5–2 keV, top) and hard (2–7 keV, bottom) X-ray emission. The centroid and radial extent of each of the four circular regions corresponding to each quasar lensed image match those in Figure~\ref{fig:chandra_data_simulations}.~Error bars in the number of counts correspond to $1\sigma$ statistical uncertainties on the number of photon counts enclosed by each region.~The color scale is logarithmic.}
    \label{fig:chandra_data_simulations_soft-hard}
\end{figure}

Figure \ref{fig:chandra_data_simulations_soft-hard} shows the number of photon counts in the soft and hard X-ray bands in the observed \chandra\ data (ObsID: 23825), showing that the soft/hard X-ray ratios are roughly conserved in images A, B, and C but \emph{not} in image D. In image D, where the soft-to-hard X-ray flux is enhanced by almost a factor of 2. Such an enhancement cannot be attributed to an additional soft X-ray component in the source, since this enhancement would otherwise be reflected in image B (the mirror image of D;~Figure \ref{fig:results_fullband}). This enhancement is therefore indicative of microlensing in image D \citep[see also][]{stern_2021_j0659+massmod,dux_2025_j659+tdels},~which must affect the distinct soft and hard X-ray emitting regions differently due to their different extents. While such an investigation is beyond the scope of this paper, we highlight that, paired with the results above, further X-ray and optical observations of J0659 could help localize the soft and hard X-ray emitting regions with greater precision, but also enable determining their relative sizes.   

\subsection{Prospects for  uncovering complex AGN morphology at high-$z$}

 \label{subsec:agn_morphology}

We highlight the prospects for undertaking broadband and spectrally resolved X-ray astrometry at milliarcsecond precision for probing quasar structure at high-$z$ in upcoming works. 

Previous X‑ray studies of lensed quasars have significantly concentrated on detecting broadened Fe K$\alpha$ features so as to infer SMBH spins in cosmologically distant AGN from X-ray reflection spectroscopy, assuming that the inner disk truncates at the innermost stable circular orbit \citep{bambi+brenneman_2021_review_spins,reynolds_2021_review_spins}. Currently, there are two lensed quasars with existing SMBH spin estimates in the literature: RX J1131-1231 \citep[RX J1131, $z_\mathrm{s} = 0.658$; see][]{reis_2014_rxj1131, chartas_2009_rxj1131} and Q 2237+305 \citep[$z_\mathrm{s} = 1.695$; see][]{m_reynolds_2014_q2237}. In addition, line shifts in the Fe K$\alpha$ complex in RX J1131 from microlensing have been used to estimate the size of the accretion disk in this source \citep[]{chartas_2009_rxj1131}. In contrast, our methodology could be used to directly probe the physical origin of the soft X-ray excess as it enables spatial constraints --or upper limits-- on the distinct localizations of the soft and hard X-ray emission regions of distant quasars. In parallel, the anomalous hard/soft X-ray flux ratio in image D in J0659 suggests microlensing sensitivity to different spatial extents of their respective emission regions. This highlights that the spatial scales our approach is able to resolve could uncover substantive X-ray emission yielding distinct broadband, soft, and hard X-ray emission arising from jets, mass outflows, or even binary/dual AGN. 

We emphasize the applicability of our methodology towards uncovering complex AGN morphology at high-$z$ for physical separations above tens of parsecs.~X-ray astrometry at milliarcsecond precision of an M87 analog (Section \ref{sec:introduction}) at $z_\mathrm{s}=2$ would enable the detection of an HST-1-like knot located $0\farcs007$ from the AGN, matching the projected separation observed in M87. The interpretation of such a feature would be further informed by our spectrally resolved astrometric analysis.~In parallel, our methodology would also permit mapping the extended Fe K$\alpha$ emission in an NGC~5768-equivalent (Section \ref{sec:introduction}) to $0\farcs26$ at $z_\mathrm{s} = 1.7$.~Our methodology could be used to search for extended soft X-ray emission at parsec scales in low-power radio galaxies, since most current studies cover the $z < 0.5$ range and rely on joint VLBI and \chandra\ imaging~\citep{giovannini_2001_pc-scale-low-pow}.

\subsection{Prospects to reveal AGN pairs at high-$z$}
\label{subsec:results_dual_agn}

\begin{figure*}[ht!]
\plotone{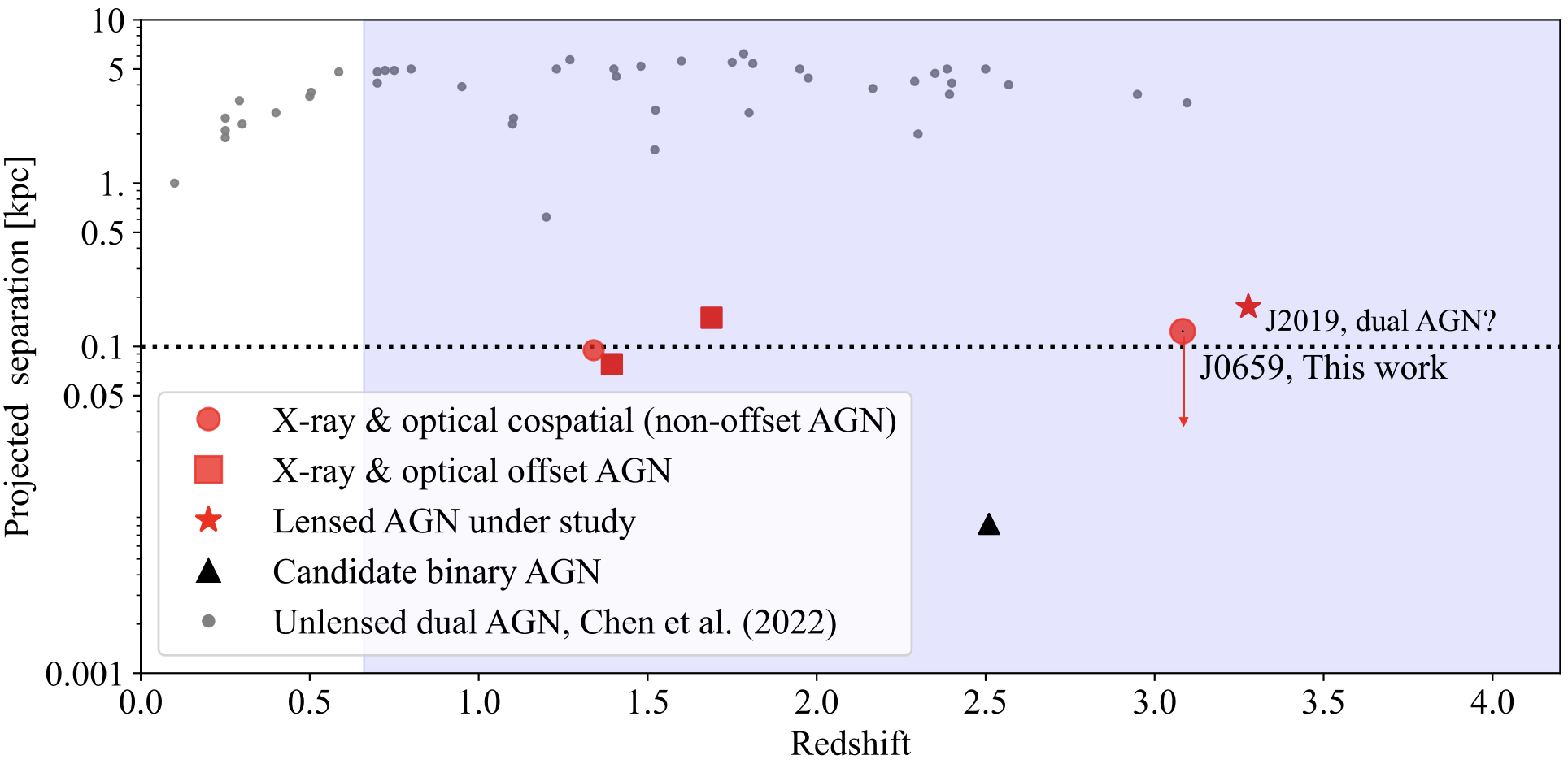}
\caption{Projected separation vs.~redshift for a sample of AGN, updated from \citet{chen_2022_varstrometry} and \citet{GLAD+spingola_2022_2systems}. The gray points correspond to the \hst-discovered unlensed pairs through optical varstrometry in \hst\ from \citet{chen_2022_varstrometry}.~The shaded region in blue outlines the discovery space for rare AGN and complex AGN morphology of strongly lensed quasars that have existing \gaia\ DR3 detections and archival \chandra\ observations.~The results from this work are highlighted, showing an upper limit in the possible X-ray-to-optical offset for J0659 at approximately 95.5\% confidence (see discussions in Sections \ref{sec:results}~and~\ref{subsec:results_fullband}).~Currently, we are undertaking a new \chandra\ analysis of the dual AGN candidate J2019~\citep{spingola_2019_j2019_vlbi,GLAD+schwartz_2021_j2019}. The projected distance of $175 \ \mathrm{pc}$ corresponds to the expected projected separation of the two putative AGN at the system redshift, $z = 3.27$ \citep{spingola_2019_j2019_vlbi}. Other quasars whose X-ray and optical offsets are plotted correspond to the offset candidates HE 0435 \citep{GLAD+rogers_2025_he0435} and CLASS B1608+656 \citep{GLAD+spingola_2022_2systems}. Similar to J0659, the radio and optical emission in CLASS B1608+656 were found to be cospatial, setting an effective upper limit on the projected separation between these emitting regions of 78 pc at the $1\sigma$ level. The projected separation between the two AGN cores in the dual SMBH candidate in the blazar PKS1830-211 is shown \citep{nair_2005_binary}. The black dotted line delimits the distinction between binary AGN and dual AGN \citep{burke-spolaor_2018_proc}.
\label{fig:chen_updated}}
\end{figure*}

Figure \ref{fig:chen_updated} illustrates how our methodology broadens the search for rarer systems, including offset, dual, and binary AGN candidates at high-$z$ beyond the capabilities of current and even future spectroscopic and imaging surveys.

Offset AGN -- possible signatures of recoiling SMBHs -- are observationally elusive but predicted outcomes of hierarchical structure formation \citep{blecha_2008_recoiling, blecha_2011_recoiling}. Separately, the search for dual and binary AGN at high redshifts \citep[pairs of AGN separated by $> 100 \ \mathrm{pc}$ and $< 100 \ \mathrm{pc}$, respectively;][]{burke-spolaor_2018_proc} also motivates the need to probe parsec‑scale separations. Detecting dual high-$z$ SMBHs at sub-kiloparsec separation provides a proxy for galaxy merger rates at early cosmic times and provides critical input for, e.g. gravitational-wave forecasts with \emph{LISA}~\citep{degraf_2024_smbhs+lisa,rantala_2025_smbhs+lisa}. Binary AGN systems are pairs of gravitationally bound SMBHs at separations $< 100 \ \mathrm{pc}$ which will eventually coalesce and yield gravitational-wave signals detectable by \lisa \citep{khusid_2023_lisa}.~Although binary AGN with parent masses $10^{8-10} M_\odot$ are believed to be the progenitors of the cosmological stochastic gravitational wave background signal detected by \textit{NANOGrav}, the identification of a single, isolated binary AGN at high confidence is yet to be made \citep[][]{dorazio+charisi_2023_SMBH-binaries}. 

We are currently undertaking a spectral analysis of a new 185 ks \chandra\ observation of the multiply imaged gravitationally lensed system J2019+1127 to test the dual AGN hypothesis in this radio loud multiply imaged system at $z_\mathrm{s}=3.28$~(\emph{Schwartz et al.,~in prep}). The confirmation this is indeed a SMBH pair at $\sim 175 \ \mathrm{pc}$ separation at its high redshift (Fig. \ref{fig:chen_updated}) would challenge the SMBH merger rate assumed by hydrodynamic simulations of cosmic structure formation within the standard paradigm of cosmology \citep[see][and references therein]{spingola_2019_j2019_vlbi,spingola_2023_nature}. 

Probing optical/radio vs.~X-ray offsets with our methodology constitutes a compelling pathway to search for evidence of AGN pairs at high redshifts. Historically, X-ray emission has been a ubiquitous indicator of AGN activity. Over recent decades, the identification of AGN pairs enabled by \chandra’s excellent imaging capabilities has been limited virtually exclusively to the local universe \citep{koss_2012_duals}. This capability has recently been expanded to search for dual AGN candidates at $z\lesssim 0.3$ with the Bayesian code \baymax.~\baymax\ is a tool that computes the Bayes factor to compare the likelihood of a single or collection of \chandra\ observations under two hypotheses (single vs. dual AGN)~and was first introduced and applied in \citet{foord_2019_baymax_i,foord_2020_baymax_ii}. However, even with future technology with smoother PSF response compared to that of \chandra, such as the Advanced X-ray Imaging Satellite (\axis), the identification of AGN pairs with \baymax\ will likely be limited to systems at projected separations $> 100 \ \mathrm{pc}$ \citep[see figure 5 of][]{foord_2024_universe+axis}, unless enhanced with strong lensing, thus underscoring the relevance of our methodology.

Figure \ref{fig:chen_updated} highlights the projected separation between the two AGN cores in the SMBH binary candidate in the blazar PKS~1830-211 \citep{nair_2005_binary}.~We anticipate that strong lensing will readily complement SMBH binary searches with \rubin\ as the latter may require monitoring on five-to-ten year baselines to reliably identify binary periods whilst mitigating noise confusion~\citep{davis_2025_SMBH-binaries+Rubin}.

\section{Conclusions}
\label{sec:conclusions}

We have performed milliarcsecond X-ray astrometry of the radio-quiet quasar GraL J0659+1609 at $z = 3.083$ by enhancing the limited spatial resolution of the \chandra\ X-ray Observatory at cosmological distances by using the caustic method -- a technique that leverages the flux magnification and the spatial amplification by strong gravitational lensing. We localize the X-ray emission in J0659 to within a \ellipsemajor$\times$\ellipseminor~ellipse at the 68.5\% level, centered $0\farcs014$ away from the optical source. We find the optical and X-ray source to be spatially coincident at $\approx 95.5\%$ confidence to within an angular offset of $0\farcs016$. At the source redshift, this sets a maximum possible projected distance between the optical and X-ray emission in J0659 of~\offset. That is, if an actual X-ray-to-optical spatial offset were present in J0659, it would correspond to a projected distance below that currently resolvable by the available~\chandra~data.~We have also demonstrated the applicability of our methodology for determining the origin of the soft and hard X-ray emitting regions to milliarcsecond accuracy in quadruply lensed quasars.~We have determined the distinct spatial origins of the soft and hard X-ray-emitting regions in J0659 to within a \ellipsemajor\ $\times$ \ellipseminor\ ellipse at the $1\sigma$ level.~The anomalous soft-to-hard flux ratio in one of the quasar lensed images suggests that microlensing is at work in this image, and that it affects the soft and hard X-ray emission regions in the source differently due to their different relative sizes. These nuances motivate the need for follow-up \chandra\ observations of J0659.~In addition to permitting milliarcsecond X-ray astrometry at high redshifts, we anticipate that strong gravitational lensing will become an important tool for the discovery of AGN pairs in current and possibly future surveys. Our methodology will also enable the study of complex AGN morphology at otherwise inaccessible redshifts in view of the thousands of quadruply lensed quasars anticipated to be discovered by \euclid, \rubin\ and \emph{Roman}. 

\begin{acknowledgments}
Support for this work was provided by the National Aeronautics and Space Administration through a NASA Astrophysics Data Analysis Program Grant 80NSSC24K0617~``Using Observations Of Gravitational Lenses To Elucidate Parsec-Scale X-Ray/Optical Structure In The Cores Of Distant Quasars’’, and by Chandra X-ray Center grants AR3-24007X and AR4-25006X. C.~S. acknowledges financial support from INAF $-$ Ricerca Fondamentale 2024 (Ob.~Fu. 1.05.24.07.04) and from the Italian Ministry of University and Research (grant FIS2023$-$01611, CUP C53C25000300001).
G.M. acknowledges financial support from INAF $-$ Ricerca Fondamentale 2022 (Ob.~Fu. 1.05.12.04.04).

This research has made use of data obtained from the \chandra\ Data Archive --including the \chandra\ Source Catalog--,~as well as software provided by the \chandra\ X-ray Center (CXC) in the application packages \ciao, \sherpa, \textsc{SAOImageDS9} (\textsc{DS9}) and the high-fidelity simulation package \saotrace.~This work has also employed data from the \gaia\ Archive (specifically, data from \gaia\ Data Releases 2 and 3) and from the \hubble\ Space Telescope (\hst) through the Barbara A. Mikulski Archive for Space Telescopes (MAST) online data archive.~The specific observations analyzed can be accessed via\dataset[doi:10.17909/m9r1-9229]{https://doi.org/10.17909/m9r1-9229}.

We thank the referee for useful comments and Martin Millon for useful remarks. JSR thanks Matthew Temple, Rafael Mart\'inez-Galarza and Aneta Siegiminowska for useful discussions in the run-up to preparing this paper.~JSR thanks Diab Jerius for valuable discussions relating to the \saotrace\ simulations we have presented. JSR thanks Yazmin Bath for contributions to the code computing the maximum likelihood.

\end{acknowledgments}

\facilities{Chandra X-ray Observatory, Gaia Telescope, Hubble Space Telescope.}

\software{\ciao~\citep{ciao-fruscione-2006}, \textsc{ds9}v8.7b1  \citep{ds9-joye-2003}, Python's \texttt{astropy} \citep{astropy-collab-2013, astropy-collab-2018, astropy-collab-22}, \texttt{matplotlib} \citep{matplotlib-release-hunter-2007}, \texttt{numpy} \citep{numpy-i-vanderwalt-2011,numpy-ii-harris-2020}, and \texttt{scipy} \citep{scipy-release-2020-virtanen} packages, \gravlens\ lensing package \citep{keeton-gravlens-2001,keeton-gravlens-2001-ii}, \saotrace\ \citep{saotrace-jerius-2004}, \sherpa v4.17.1 \citep{sherpa-release-freeman-2001,sherpa-v4.17.1-burke-2024}, and \xspec\ v12.14.1 \citep{arnaud-xspec-1996}. }

\bibliography{sample701}

\end{document}